# New Quasi-Cyclic Codes from Simplex Codes

Eric Z. Chen

*Abstract*— As a generalization of cyclic codes, quasi-cyclic (QC) codes contain many good linear codes. But quasi-cyclic codes studied so far are mainly limited to one generator (1-generator) QC codes. In this correspondence, 2-generator and 3-generator QC codes are studied, and many good, new QC codes are constructed from simplex codes. Some new binary QC codes or related codes, that improve the bounds on maximum minimum distance for binary linear codes are constructed. They are 5-generator QC [93, 17, 34] and [254, 23, 102] codes, and related [96, 17, 36], [256, 23, 104] codes.

## I. INTRODUCTION

A code is said to be quasi-cyclic if every cyclic shift of a codeword by $p$ positions results in another codeword. Therefore quasi-cyclic (QC) codes are a generalization of cyclic codes with $p = 1$. It has been shown that QC codes contain many good linear codes [1-4]. Unfortunately, there are not many construction methods for good QC codes. Lots of researchers have turned to the power of modern computers, and many good QC codes which improve lower bounds on the minimum distance of linear codes have been found [5-8]. The author maintains a database of best-known binary QC codes [9].

In [15], two 2-generator QC codes were presented. Very little research on how to construct good g-generator QC codes has been made, with g > 1. In this correspondence, construction methods are presented to construct good 2-, and 3-generator QC codes from cyclic simplex codes. Many good, new QC codes are found. And several codes that improve the bounds on maximum minimum distance for binary linear codes are also presented.

## II. CYCLIC CODES AND QC CODES

### A. Cyclic Hamming Codes and Simplex Codes

A q-ary linear [n, k, d] code is a k-dimensional subspace of an n-dimensional vector space over GF(q), with minimum distance d between any two codewords. A code is said to be cyclic if every cyclic shift of a codeword is also a codeword. A cyclic code is described by the polynomial algebra. A cyclic [n, k, d] code has a unique generator polynomial g(x). It is a polynomial with degree of n – k. All codewords of a cyclic code are multiples of g(x) modulo $x^n - 1$.

Eric Z. Chen is with School of Engineering, Kristianstad University, 291 88 Kristianstad, Sweden( eric.chen@tec.hkr.se).

It is well known that for any positive integer k, there is a binary cyclic simplex [n, k, d] code with the minimum distance $d = 2^{k-1}$, where $n = 2^k - 1$. It should be noted that binary simplex codes are equidistance codes where $2^k - 1$ non-zero codewords have a weight of $2^{k-1}$.

### B. Quasi-Cyclic Codes

A code is said to be quasi-cyclic (QC) if a cyclic shift of any codeword by p positions is still a codeword. Thus a cyclic code is a QC code with $p = 1$. The block length n of a QC code is a multiple of $p$, or $n = m \times p$.

Circulants, or cyclic matrices, are basic components in the generator matrix for a QC code. An $m \times m$ cyclic or circulant matrix is defined as

$$C = \begin{bmatrix} c_0 & c_1 & \cdots & c_{m-1} \\ c_{m-1} & c_0 & \cdots & c_{m-2} \\ c_{m-2} & c_{m-1} & \cdots & c_{m-3} \\ \vdots & \vdots & \cdots & \vdots \\ c_1 & c_2 & \cdots & c_0 \end{bmatrix} \quad (1)$$

and it is uniquely specified by a polynomial formed by the elements of its first row, $c(x) = c_0 + c_1 x + \ldots + c_{m-1} x^{m-1}$, with the least significant coefficient on the left.

A 1-generator QC code has the following form of the generator matrix [11]:

$$G = [\, G_0 \; G_1 \; G_2 \; \ldots \; G_{p-1} \,] \quad (2)$$

where $G_i, i = 0, 1, 2, \ldots, p-1$, are circulants of order m. Let $g_0(x), g_1(x), \ldots, g_{p-1}(x)$ are the corresponding defining polynomials.

A 2-generator QC [$m \times p, k$] codes has the generator matrix of the following form:

$$G = \begin{bmatrix} G_{00} & G_{01} & \cdots & G_{0,p-1} \\ G_{10} & G_{11} & \cdots & G_{1,p-1} \end{bmatrix} \quad (3)$$

where $G_{ij}$ are circular matrices, for $i = 0$, and $1$, $j = 0, 1, \ldots, p-1$.

Similarly, a 3-generator QC [$m \times p, k$] codes has the generator matrix of the following form:

$$G = \begin{bmatrix} G_{00} & G_{01} & \cdots & G_{0,p-1} \\ G_{10} & G_{11} & \cdots & G_{1,p-1} \\ G_{20} & G_{21} & \cdots & G_{2,p-1} \end{bmatrix} \quad (4)$$

where $G_{ij}$ are circular matrices, for i = 0, 1, and 2, j = 0, 1, …, p-1.

## III. CONSTRUCTIONS OF 2-GENERATOR QC CODES

Given any positive integer k. If there exist two binary cyclic Hamming $[2^k – 1, 2^k – k – 1, 3]$ codes, then there exist two cyclic simplex $[2^k – 1, k, 2^{k-1}]$ codes. Let $g_1(x)$ and $g_2(x)$ be the generator polynomials of these simplex codes, $C_1$ and $C_2$. A binary 2-generator QC $[(2^k – 1)p, 2k]$ code can be constructed with the following generator matrix:

$$G = \begin{bmatrix} G_1 & G_1 & ... & G_1 \\ G_2 & G_{2,1} & ... & G_{2,p-1} \end{bmatrix} \quad (5)$$

where $G_1$ is the circulant matrix defined by the generator polynomial $g_1(x)$, $G_2$ is the circulant matrix defined by $g_2(x)$, and $G_{2,i}$ is the circulant matrix defined by $x^{a(i)}g_2(x)$, where $0 \leq a(i) < 2^k – 1$, is an integer. The choices of a(i), i = 1, 2, …, p – 1, are to maximize the minimum distance to the code.

Let $m = 2^k – 1$, and the distance vector $D = (d_0, d_1, …, d_{m-1})$, where $d_i$ is defined as the distance between the codeword $g_1(x)$ in $C_1$ and the codeword $x^i g_2(x)$ in $C_2$. Then a good 2-generator QC [m×p, 2k] code can be obtained by choosing a(i) to maximize its minimum distance:

$d = \min ( d_j + d_{j+a(1)} + … + d_{j+a(p-1)} )$,
where j = 0, 1, …, m – 1.

Example 1. n = 7, k = 3. $x^7 – 1 = (x + 1)(x^3 + x + 1)(x^3 + x^2 +1)$. So two cyclic simplex [7, 3, 4] codes are defined repectively by $g_1(x) = x^4 + x^3 + x^2 + 1$ and $g_2(x) = x^4 + x^2 + x + 1$. The distance vector D = (2, 4, 4, 6, 2, 2, 4). Let p = 3, an optimal 2-generator QC [21, 6, 8] code can be obtained with a(1) = 1, and a(2) = 2.

For n = 15, k = 4. $x^{15} – 1 = (x + 1)(x^2 + x+ 1)(x^4 + x + 1)(x^4 + x^3 + 1)(x^4 + x^3 + x^2 +x + 1)$. Two cyclic simplex [15, 4, 8] codes can are defined repectively by $g_1(x)$ = 7531 and $g_2(x)$ = 4657 in octal, with the highest degree terms on the left. The distance vector is D = (8, 6, 8, 4, 6, 10, 8, 6, 4, 10, 6, 10, 10, 8, 8).

For n = 31, k = 5. Two cyclic simplex [31, 5, 16] codes are defined respectively by $g_1(x)$ = 454761565 and $g_2(x)$ = 715750453 in octal. The distance vector is D = (12, 12, 16, 16, 20, 12, 16, 20, 16, 16, 20, 16, 16, 16, 16, 12, 16, 20, 16, 12, 12, 12, 16, 12, 20, 16, 16, 16, 12, 20, 12).

For n = 63, k = 6. Two cyclic simplex [63, 6, 32] codes are defined respectively by $g_1(x)$ = 10305172162267315277 and $g_2(x)$ = 13745214756551542207 in octal. The distance vector is D = (32, 32, 24, 40, 32, 32, 40, 32, 32, 32, 32, 32, 24, 32, 32, 32, 32, 24, 32, 40, 32, 32, 32, 32, 32, 40, 32, 32, 32, 32, 24, 32, 32, 32, 32, 32, 32, 32, 24, 24, 24, 32, 24, 40, 32, 32, 24, 32, 40, 32, 32, 32).

Table I lists good binary 2-generator QC codes constructed. In the table, superscript "o" denotes the code obtained is optimal, "=" denotes the code meets the best minimum distance in [12].

Many projective two-weight codes are also constructed. For examples, binary 2-generator QC two-weight codes with p = 3, 4, 7 for m = 7, and p = 10, 11, 12, and 15 for m = 15 are obtained. These codes have the same parameters as the family SU2 [13]. In [14], quasi-cyclic QC two-weight codes are discussed and lot of codes are constructed.

Table I GOOD BINARY 2-GENERATOR QC CODES

| p | n | k | d | a(i) |
|---|---|---|---|------|
| 3 | 7 | 3 | 8° | 1, 2 |
| 4 | 7 | 3 | 12° | 1, 2, 4 |
| 5 | 7 | 3 | 16° | 0, 3, 5, 6 |
| 7 | 7 | 3 | 24° | 1, 2, 3, 4, 5, 6 |
| 2 | 15 | 4 | 12° | 3 |
| 5 | 15 | 4 | 34° | 1, 3, 6, 13 |
| 10 | 15 | 4 | 72° | 1, 3, 4, 6, 7, 9, 10, 12, 13 |
| 11 | 15 | 4 | 80° | 1, 2, 3, 4, 5, 6, 9, 11, 12, 13 |
| 12 | 15 | 4 | 88° | 1, 2, 3, 5, 6, 7, 8, 10, 11, 12, 13 |
| 15 | 15 | 4 | 112° | 1, 2, 3, 4, 5, 6, 7, 8, 9, 10, 11, 12, 13, 14 |
| 5 | 31 | 5 | 72= | 1, 3, 6, 18 |
| 6 | 31 | 5 | 88= | 1, 4, 7, 19, 28 |
| 2 | 63 | 6 | 56= | 21 |
| 3 | 63 | 6 | 88° | 21, 42 |

## IV. CONSTRUCTIONS OF 3-GENERATOR QC CODES

Similarly, if there exist 3 binary cyclic simplex $[2^k – 1, k, 2^{k-1}]$ codes, $C_1$, $C_2$, and $C_3$, defined by generator polynomials $g_1(x)$, $g_2(x)$, and $g_3(x)$, respectively, then a binary 3-generator QC $[(2^k – 1)p, 3k]$ code can be constructed as follows:

$$G = \begin{bmatrix} G_1 & G_1 & ... & G_1 \\ G_2 & G_{2,1} & ... & G_{2,p-1} \\ G_3 & G_{3,1} & ... & G_{3,p-1} \end{bmatrix} \quad (6)$$

where $G_1$ is the circulant matrix defined by the generator polynomial $g_1(x)$, $G_2$ is the circulant matrix defined by $g_2(x)$, $G_3$ is the circulant matrix defined by $g_3(x)$, $G_{2,i}$ is the circulant defined by $x^{a(i)}g_2(x)$, $G_{3,i}$ is the circulant defined by $x^{b(i)}g_3(x)$, where $0 \leq a(i) < 2^k – 1$, and $0 \leq b(i) < 2^k – 1$. The choices of integers a(i) and b(i), i = 1, 2, …, p – 1, are to maximize the minimum distance to the code.

Let $m = 2^k – 1$, and the distance vectors $D_{12} = (d_0^{12}, d_1^{12}, …, d_{m-1}^{12})$, $D_{13} = (d_0^{13}, d_1^{13}, …, d_{m-1}^{13})$, $D_{23} = (d_0^{23}, d_1^{23}, …, d_{m-1}^{23})$, where $d_i^{12}$ is the distance between the codeword $g_1(x)$ in $C_1$ and the codeword $x^i g_2(x)$ in $C_2$, $d_i^{13}$, the distance between the codeword $g_1(x)$ in $C_1$ and the codeword $x^i g_3(x)$ in $C_3$, and $d_i^{23}$ the distance between the codeword $g_2(x)$ in $C_2$ and the codeword $x^i g_3(x)$ in $C_3$. Let the distance table $D_{123}$ be

$$d_{0,0}^{123}, d_{0,1}^{123}, ..., d_{0,m-1}^{123}$$
$$d_{1,0}^{123}, d_{1,1}^{123}, ..., d_{1,m-1}^{123}$$
$$...$$
$$d_{m-1,0}^{123}, d_{m-1,1}^{123}, ..., d_{m-1,m-1}^{123}$$

where $d_{i,j}^{123}$ is the weight of the sum of codewords $g_1(x)$ in $C_1$, $x^i g_2(x)$ in $C_2$ and $x^j g_3(x)$ in $C_3$.

Then a good 3-generator QC [m×p, 3k] code can be obtained by choosing integers a(i) and b(j) to maximize its minimum distance:

$$d_j^{12} + d_{j+a(1)}^{12} + ... + d_{j+a(p-1)}^{12},$$
$$d_j^{13} + d_{j+b(1)}^{13} + ... + d_{j+b(p-1)}^{13},$$
$$d_j^{23} + d_{j+b(1)-a(1)}^{23} + ... + d_{j+b(p-1)-a(p-1)}^{23},$$
$$d_{i,j}^{123} + d_{i+a(1),j+b(1)}^{123} + ... + d_{i+a(p-1),j+b(p-1)}^{123}$$

where i, j = 0, 1, ..., m – 1, and subscripts are computed modulo m.

For k = 5, three cyclic simplex [31, 5, 16] codes are defined by generator polynomials $g_1(x)$ = 535437151, $g_2(x)$ = 454761565 and $g_3(x)$ = 715750453 in octal. For p = 3, a 3-generator QC [93, 15, 36] code is obtained with a(1) = 1, a(2) = 18, and b(1) = 30, b(2) = 2. This code meets the lower bound on the minimum distance. For the code constructed in the methods given above, It is often possible to extend the code by adding one or more information digits and parity check digits. Based on the [93, 15, 36] code, a new 5-generator QC [93, 17, 34] code is obtained. Its generator matrix is shown below:

$$G = \begin{bmatrix} G_1 & G_1 & G_1 \\ G_2 & G_{2,1} & G_{2,2} \\ G_3 & G_{3,1} & G_{3,1} \\ 1...1 & 1...1 & 0...0 \\ 1...1 & 0...0 & 1...1 \end{bmatrix}$$

Further, if three additional parity check digits(one for each 31 bits) are added, a new binary [96, 17, 36] code is obtained. Both [93, 17, 34] and [96, 17, 36] codes improve the lower bound on the minimum distance on binary linear codes.

For k = 7, three cyclic simplex [127, 7, 64] codes are defined by generator polynomials $g_1(x)$ = 001772514735130 6755331107027625632117050301, $g_2(x)$ = 11151734177073 051372502674712630155350621 and $g_3(x)$ = 141117737072 51275147153042731036267012155 in octal. A 3-generator QC [254, 21, 104] code can be obtained with a(1) = 21, and b(1) = 43. An additional information digit can be added to this code by adding two extra rows to the generator matrix, a new 5-generator QC [254, 23, 102] code can be obtained with the following generator matrix:

$$G = \begin{bmatrix} G_1 & G_1 \\ G_2 & G_2 \\ G_3 & G_3 \\ 1...1 & 0...0 \\ 0...0 & 1...1 \end{bmatrix}$$

If two additional parity check digits(one for each 127 bits) are added, a new linear [256, 23, 104] code can be obtained. Both [254, 23, 102], and [256, 23, 104] codes improve the lower bound on the minimum distance on binary linear code.

## V. CONCLUSION

In this correspondence, construction methods for 2- and 3-generator QC codes from cyclic simplex codes are presented. Many good new 2-, 3-generator QC codes are found, and new linear codes that improve the lower bounds on the minimum distance are constructed.

REFERENCES

[1] C. L. Chen and W.W. Peterson, "Some results on quasi-cyclic codes", Infom. Contr., vol. 15, pp.407-423, 1969.
[2] E. J. Weldon, Jr., "Long quasi-cyclic codes are good", IEEE Trans. Inform. Theory, vol.13,no.1, p.130, Jan. 1970.
[3] T. Kasami, "A Gilbert-Varshamov bound for quasi-cyclic codes of rate 1/2", IEEE Trans. Inform. Theory, vol. IT-20, p.679, 1974.
[4] San Ling and Patrick Solé, "Good Self-Dual Quasi-Cyclic Codes Exist", IEEE Trans. Inform. Theory, vol.39, pp.1052-1053, 2003.
[5] H.C.A. van Tilborg, "on quasi-cyclic codes with rate 1/m", IEEE Trans. Inform. Theory, vol.IT-24, no.5, pp.628-629, Sept. 1978.
[6] T.A. Gulliver and V.K. Bhargava, "Some best rate 1/p and rate (p-1)/p systematic quasi-cyclic codes", IEEE Trans. Inform.Theory, vol.IT-37, no.3, pp.552-555, May 1991.
[7] Eric Zhi Chen, "Six new binary quasi-cyclic codes", IEEE Trans. Inform. Theory, vol.IT-40, no.5, pp.1666-1667, Sept. 1994.
[8] Petra Heijnen, Henk van Tilborg, Tom Verhoeff, and Sander Weijs, "Some new binary quasi-cyclic codes", IEEE Trans. Inform. Theory, vol. 44, 1994-1996, Sept. 1998.
[9] Eric Zhi Chen, Web database of binary QC codes, [Online]. Available: http://www.tec.hkr.se/~chen/research/codes/searchqc2.htm
[10] F. J. MacWilliams and N.J.A. Sloane, The theory of error-correcting codes, North Holand, Amsterdam, 1977.
[11] G. E. Séguin and G. Drolet, "The theory of 1-generator quasi-cyclic codes", manuscript, Dept of Electr. and Comp. Eng., Royal Military College of Canada, Kingston, Ontario, June 1990.
[12] A. E. Brouwer, "Bounds on the minimum distance of linear codes ", [Online]. Available: http://www.win.tue.nl/~aeb/voorlincod.html.
[13] R. Calderbank and W. M. Kantor, "The geometry of two-weight codes", Bull. London Math. Soc., vol. 18, pp.97—122, 1986.
[14] Eric Z. Chen, "Constructions of Quasi-Cyclic Two-weights Codes", Proc. of the 10th Int. Workshop on Algebraic and Combinatorial Coding Theory, *Zvenigorod, Russia, September 03-09, 2006*.
[15] T.A. Gulliver and V.K. Bhargava, "Two new rate 2/p binary quasi-cyclic codes", IEEE Trans. Inform. Theory, Vol. 40, pp.1667-1668, Sept. 1994